\documentstyle[sprocl,psfig]{article}

\def\ra{\rightarrow}

\begin{document}
\rightline{\vbox{\baselineskip=14pt
\rightline{UH-511-955-00} \break
  \rightline{February 2000}
}}
\vspace{.25in}
\title{CP VIOLATION IN HYPERON DECAYS}

\author{S. PAKVASA}

\address{Department of Physics \\
 University of Hawaii at Manoa \\
            Honolulu, HI 96822  \\ Email:  pakvasa@phys.hawaii.edu}

\maketitle\abstracts{The theory and phenomenology of CP violation 
in hyperon decays is summarized.}

\section{Introduction}

The CPT theorem was proved in 1955\cite{luders} and soon thereafter 
L$\ddot{\rm{u}}$ders and 
Zumino\cite{luderss} deduced from it the equality of masses and lifetimes 
between particles and
anti-particles. In 1958 Okubo\cite{okubo} observed that CP violation allows 
hyperons and
antihyperons to have different branching ratios into conjugate channels
even though their total rates must be equal by CPT. Somewhat later, this paper
inspired Sakharov\cite{sakha} to his famous work on cosmological baryon-antibaryon
asymmetry.  In fact, he called this the ``Okubo effect'', perhaps a better phrase than
the current dull use of ``direct CP violation.''
Pais\cite{pais} extended Okubo's proposal to asymmetry parameters in
$\Lambda$ and $\bar{\Lambda}$ decays. The subject was revived in the
'80s and a number of calculations were made\cite{brown,chau}.  Only now, 
over 40 years after Okubo's paper, are 
these proposals about to be tested in the laboratory.

The reason for the current interest is the need to find CP violation in
places other than just $K_L-K_S$ complex.  Only a number of different
observations of CP violation in different channels will help us pin down
the source and nature of CP violation in or beyond the standard model (SM).
From this point of view, hyperon decay is one more weapon in our arsenal
in addition to
the K system, the B system, the D system,  etc.

\section{Phenomenology of Hyperon Decays}

I summarize here the salient features of the phenomenology of
non-leptonic hyperon decays \cite{rosen}.  Leaving out $\Omega^-$ decays, there  are
seven decay modes $\Lambda \ra N \pi, \ \Sigma^\pm \ra N \pi$ and $\Xi \ra \Lambda \pi$.  The
effective matrix element can be written as 
\begin{equation}
i \ \bar{u}_{\bar{p}} (a + b \gamma_5) u_\Lambda \ \phi
\end{equation}
for the mode $\Lambda \ra p + \pi^-$, where a and b are complex in general.
The corresponding element for $\bar{\Lambda} \ra \bar{p} + \pi^+$ is
then:
\begin{equation}
i \ \bar{v}_{\bar{p}} (-a^* + b^* \gamma_5) v_{\bar{\Lambda}} \phi^+
\end{equation}
It is convenient to express the observables in terms of S and  P and
write the matrix element as
\begin{equation}
S + P \ {\bf \sigma}.\hat{\bf q}
\end{equation}
where {\bf q} is the proton momentum in the $\Lambda$ rest frame
and S and P are:
\begin{eqnarray}
S &=& a \sqrt{ \left\{ (m_\Lambda + m_p)^2 - m^2_\pi \right\}
\over 
16 \pi \ m^2_{\Lambda}} \nonumber \\
P &=& b \sqrt {\left\{ (m_\Lambda - m_p)^2 - m^2_\pi \right\}
\over 16 \pi \ m^2_{\Lambda}}
\end{eqnarray}
In the $\Lambda$ rest-frame, the decay distribution is given by:
\begin{eqnarray}
\frac{d \Gamma}{d \Omega} &=&
\frac{\Gamma}{8 \pi} %
\{ 
[1 + \alpha < {\bf \sigma}_\Lambda > . \hat{\bf \sigma} ] \nonumber \\
&+& < {\bf \sigma}_p >. [( \alpha + < {\bf \sigma}_\Lambda > . 
\hat{\bf q}) \hat{\bf q}+ \ \ \beta < {\bf \sigma}_\Lambda > \times \hat{\bf q} \nonumber \\
&+& \gamma ( \hat{\bf q} \times ( < {\bf \sigma}_\Lambda > \times
 \hat{\bf q})] 
 \}
\end{eqnarray}  
where $\Gamma$ is the decay rate and is given by:
\begin{equation}
\Gamma = 2 \mid {\bf q} \mid \left\{ \mid S \mid^2 + \mid P \mid^2 \right \}
\end{equation}
$\alpha, \beta$ and $\gamma$ are given by
\begin{eqnarray}
\alpha &=& 2Re (S^*P ) \over \left \{ \mid S \mid^2 + \mid P \mid^2 \right
\}, \nonumber \\
\beta &=& 2 Im (SP^*) \over \left \{ \mid S \mid^2 + \mid P \mid^2
\right \}
\nonumber \\
\gamma &=&  \left \{ \mid S \mid^2 - \mid P \mid^2 \right \} \over 
\left \{ \mid S \mid^2 + \mid P \mid^2 \right \}
\end{eqnarray}
For a polarized $\Lambda$, the up-down asymmetry of the final proton is
given by  $\alpha ( \alpha$ is also the longitudinal polarization of the
proton for an unpolarized $\Lambda)$.  $\beta$ and $\gamma$ are
components of the transverse polarization of proton \cite{lee}.

The observed properties of the hyperon decays can be summarised as:  (i)
the $\Delta I = 1/2$ dominance i.e. the $\Delta I = 3/2$ amplitudes are about
5\% of the $\Delta I = 1/2$ amplitudes; (ii) the asymmetry parameter
$\alpha$ is large for $\Lambda$ and $\Xi$ decays, $\Xi$ decays and $\Sigma^+ \ra p
\pi^0$ and is near zero for $\Sigma^\pm \ra n \pi^\pm$; and (iii) the
Sugawara-Lee triangle sum rule $\sqrt{3}A ( \Sigma^+ \ra p  \pi^0) - A 
(\Lambda \ra p \pi^-) = 2A (\Xi \ra \Lambda \pi^-)$ is satisfied to a level
of 5\% in both $s$ and $p$ wave amplitudes.

\section{CP Violating Observables}

Let a particle P decay into several final states $f_1, f_2$ etc.  The
amplitude for P $\ra f_1$ is in general:
\begin{equation}
A= A_1 e^{i \delta 1} \ \ + A_2 \ e^{i \delta 2}
\end{equation}
where 1 and 2 are strong interaction eigenstates and $\delta_i$ are
corresponding final state phases.  Then the amplitude for
$\bar{P} \ra \bar{f}_1$ is
\begin{equation}
\bar{A} = A^*_1 e^{i \delta 1} \ \ + A^*_2 \ e^{i \delta 2}
\end{equation}
If $\mid A_1 \mid >> \mid A_2 \mid$, then the rate asymmetry
$\Delta ( = (\Gamma - \bar{\Gamma}) / (\Gamma + \bar{\Gamma}) )$ is
given by:
\begin{equation}
\Delta \approx -2 \mid A_2/A_1 \mid sin ( \phi_1 - \phi_2) sin(\delta_1- 
\delta_2)
\end{equation}
where $A_i = \mid A_i \mid e^{i \phi_i}$.  Hence, to get a non-zero rate
asymmetry, one must have (i) at least two channels in the final state,
(ii) CPV weak phases must be different in the two channels, and (iii) 
\underline{unequal} final state scattering phase shifts in the two
channels\cite{brown}. 
A similar calculation of the asymmetry of $\alpha$\cite{overseth} shows that
for a single isospin channel dominance,
\begin{equation}
A= (\alpha + \bar{\alpha})/ (\alpha - \bar{\alpha}) = 2 tan ( \delta_s -
\delta_p) \ tan (\phi_s - \phi_p)
\end{equation}
In this case the two channels are orbital angular momenta $0$ and $1$;
and even a single isospin mode such as $\Xi^- \ra \Lambda \pi^-$ can
exhibit a non-zero A. In B decays an example of a single isospin mode
exhibiting CP violating rate asymmetry is $B \ra \pi \pi$, i.e. 
In this case the two eigen-channels with different weak CP phases and
different final state phases are $B \ra D \overline{D} \pi \ra \pi\pi$
and $B \ra \pi \pi \ra \pi \pi$\cite{wolfenstein}. 

To define the complete set of CP violating observables, consider the
example of the decay modes $\Lambda \ra p \pi^-$ and
$\bar{\Lambda} \ra \bar{p} \pi^+$.  The amplitudes are:
\begin{eqnarray}
S&=& - \sqrt{2 \over 3} S_1 e^{i ( \delta_1 + \phi_1^s)} +
    \frac{1} {\sqrt{3}} S_3 e^{i ( \delta_3 + \phi^s_3)} \nonumber \\
P&=& - \sqrt{2 \over 3} P_1 e^{i ( \delta_{11} + \phi_1^p)} +
    \frac{1}{\sqrt{3}} P_3 e^{i ( \delta_3 + \phi^p_3)} 
\end{eqnarray}
where $S_i, P_i$ are real, $i$ refers to the final state isospin (i=2I)
and $\phi_i$ are the CPV phases.  With the knowledge that $S_3/S_1$,
$P_3/P_1 <<$ 1 ;  one can write\cite{donog,donog1}
\begin{eqnarray}
\Delta_\Lambda &=& \frac{(\Gamma-\overline{\Gamma})}{(\Gamma + \overline{\Gamma})}
\cong \sqrt{2} \ (S_3/S_1) sin ( \delta_3 - \delta_1) sin
(\phi_3^s - \phi_1^s) \nonumber \\
A_\Lambda &=& \frac{(\alpha +\overline{\alpha})} {(\alpha - \overline{\alpha})}
\cong - tan (\delta_{11} - \delta_1) tan (\phi_1^p - \phi_1^s)
\nonumber \\
B_\Lambda &=& \frac{(\beta +\overline{\beta})}{(\beta - \overline{\beta})}
\cong  cot  (\delta_{11} - \delta_1) tan (\phi_1^p - \phi_1^s)
\end{eqnarray}
The last one $B_\Lambda$ has the peculiar feature that it blows up as the phase
shift difference vanishes.  The reason is that in the limit of CP
conservation $\beta + \overline{\beta} = 0$ but in the limit of no final state
phase difference $\beta - \bar{\beta} = 0$. 
For $\pi$N final states, the phase shifts at $E_{c.m.} = m_\Lambda$ are
known and are: $\delta_1 = 6^0, \ \delta_3 = -3.8^0, \ \delta_{11} = 1.1^0$
and $\delta_{31} = -0.7^0$ from the 1965 analysis of Roper et
al.\cite{roper} 
with errors
estimated at 10\%. The CPV phases $\phi_i$ have to be provided
by theory.

Similar expressions can be written for other hyperon decays.  For
example, for $\Lambda \ra n \pi^0$, $\Delta$ is $- 2 \Delta_\Lambda$
and $A$ and $B$ are identical to $A_\Lambda$ and $B_\Lambda$.  For
$\Xi^- \ra \Lambda \pi^-$ ( and $\Xi^0 \ra \Lambda \pi^0)$ the 
asymmetries are\cite{donog1}:
\begin{eqnarray}
\Delta_\Xi & = & 0    \nonumber \\
A_\Xi & = &-tan (\delta_{21} - \delta_2) tan (\phi^p - \phi^s)
\nonumber \\
B_\Xi & = & cot (\delta_{21} - \delta_2) tan (\phi^p- \phi^s)
\end{eqnarray}
where $\delta_{21}$ and $\delta_2$ are the $p$ and $s$-wave $\Lambda
\pi$ phase shifts at $m_\Xi$ respectively.  Somewhat more
complicated expressions can be and have been written for $\Sigma$ decays\cite{donog1}.

\section{Calculating CP Phases}

In standard model description of the non-leptonic hyperon decays, the
effective $\Delta S = 1$ Hamiltonian is
\begin{equation}
H_{eff} = \frac{G_F}{\sqrt{2}} \ U_{ud}^* \ U_{us} \sum^{12}_{i=1} c_i
(\mu) \ O_i ( \mu)
\end{equation}
after the short distance QCD corrections  (LLO + NLLO) where
$c_i = z_i + y_i \tau ( \tau = -U_{td} \ U^*_{ts} / U_{ud} \ U_{us})$,
and $\mu \sim 0(1$ GeV)\cite{buras}. 
For CP violation, the most important operator is:
\begin{equation}
O_6 = \bar{d} \ \lambda_i\gamma_\mu (1+ \gamma_5) s 
\bar{q} \lambda_i \gamma_\mu (1 -\gamma_5)q
\end{equation}
and $y_6 \approx -0.1$ at $\mu \sim 1 \ GeV$.  
To estimate the CP phases in Eq. (12), one adopts the following
procedure. The real parts(in the approximation that the imaginary 
parts are very small) are known from the data on rates and asymmetries.
The real parts of the amplitudes have also been evaluated in SM with
reasonable success with some use of chiral perturbation theory(current
algebra and soft pion theorems) and a variety of choices for the 
baryonic wave functions. The MIT bag model wave function is one such
choice which gives conservative results.
The same procedure is adopted for calculating the imaginary parts using 
$O_6$. The major 
uncertainty is in the hadronic matrix elements and the fact that the 
simultaneous fit of $s$ and $p$ waves leaves a factor of 2 ambiguity
\cite{donog2}.
In the SM,
with the Kobayashi-Maskawa phase convention\cite{kobayashi} there is no CPV in $\Delta I
= 3/2$ amplitudes; and for $\Lambda$ and $\Xi$ decays $\phi_3 = 0$. There is a small
electroweak penguin contribution to $\phi_3$ which can be safely neglected. The
rate asymmetry is dominated by the s wave amplitudes and the asymmetry $A_\Lambda$
is dominated by the $\Delta I = 1/2$ amplitudes. Evaluating the 
matrix elements in the standard way and with the current knowledge of the 
K-M matrix elements one finds for the decays\cite{donog1,iqbal} $\Lambda \ra p \pi^-$ and $\Xi^-
 \ra \Lambda \pi^-$:
\begin{eqnarray}
& &\phi^s_\Lambda - \phi^p_\Lambda \cong 3.5.10^{-4}  \nonumber \\
& &\phi_\Xi^s - \phi^p_\Xi \cong - 1.4.10^{-4}
\end{eqnarray}
With the $N \pi$  phase shifts known to be 
\begin{equation}
\delta_s - \delta_p \cong 7^0
\end{equation}
one finds for the asymmetry $A_\Lambda$ in the standard model a value of 
about $-4.10^{-5}$.   For the $\Xi \ra \Lambda \pi^-$ decay mode the phase 
shifts are not known experimentally and have to be determined 
theoretically.  There are calculations from 1965 \cite{martin} which gave large values for 
$\delta_s - \delta_p \sim -20^0$; however, all recent calculations based on chiral 
perturbation theory, heavy baryon approximation etc. agree that $\delta_s - 
\delta_p$ lies between $1^0$ and $3^0$ \cite{lu}.  These techniques have
been tested in $\pi$-N scattering where they reproduce the known phase
shifts within a factor of two\cite{datta}.  In this case the asymmetry 
$A_\Xi$ is expected to be $\sim - (0.2$ to $0.7) 10^{-5}$. In the Table 1,
the SM results for the expected asymmetries in SM are given. Using very crude
back of the envelope estimates, similar results are obtained.  What is
needed is some attention to these matrix elements from the Lattice community. 

An experimental measurement of the phase shifts $\delta_s - \delta_p$ in 
$\Lambda \pi$ system will put the predictions for $A_\Xi$ on a firmer basis.  
There is an old proposal due to Pais and Treiman \cite{pais1} to measure $\Lambda \pi$ 
phase shifts in $\Xi \ra \Lambda \pi e\nu$, but this does not seem
practical in the near future.  Another technique, more feasible, it to measure $\beta$ and 
$\alpha$ to high precision in $\Xi$ and $\overline{\Xi}$ decays.  Then the 
combination.
\begin{equation}
( \beta - \bar{\beta}) / (\alpha - \bar{\alpha}) = tan \ (\delta_s -
\delta_p)
\end{equation}
can be used to extract $\delta_s -\delta_p$.  To the extend CP phases are 
negligible one can also use the approximate relation:
\begin{equation}
\beta/\alpha \approx tan (\delta_s -\delta_p)
\end{equation}

In $\Sigma$ decays, some asymmetries are quite large\cite{donog1} but in
difficult to measure channels e.g. $B_\Sigma$.  
In $\Omega^- \ra \Xi \pi$ decays the
rate asymmetry is larger due to the larger $\Delta I = 3/2$
amplitudes\cite{tandean}.  There are no experimental proposals to measure CP
asymmetries in $\Sigma$ or $\Omega^-$ decays at this time.

\section{Beyond Standard Model}

Can new physics scenarios in which the source of CP violation is not K-M 
matrix yield large enhancements of these asymmetries?   We consider some 
classes of models where these asymmetries can be estimated more or less
reliably 
\cite{donog,donog1}.  It should be kept in mind that any estimates with
new physics are at least as uncertain as SM and usually much more prone
to uncertainty for obvious reasons.

First there is the class of models which are effectively super-weak \cite{some}.
Examples are models in which the K-M matrix is real and the observed CP
violation is due to exchange of heavier particles; heavy scalars with FCNC,
heavy quarks etc.  In all such models direct CP violation is negligible and
unobservable and so all asymmetries in hyperon decays are essentially zero.
Furthermore, they need to be modified to accommodate the fact that
direct CP violation (``Okubo effect'') has now been seen in the kaon decays( the fact that
$\epsilon'/\epsilon$ is not zero).
In the three Higgs doublet model with flavor conservation imposed, the 
charged Higgs exchange tends to give large effects in direct CP violation as 
well as large neutron electric dipole moment \cite{wein}.

There are  two generic classes of left-right symmetric models:  (i) Manifest Left-
Right symmetric model without $W_L - W_R$ mixing \cite{mohap} and (ii) with $W_L -
W_R$ mixing \cite{chang}.  In (i) $U_{KM}^L =$ real and $U_{KM}^R$ complex 
with arbitrary phases but angles given by $U_{KM}^L$.  Then one gets 
the ``isoconjugate'' version in which
\begin{equation}
H_{eff} = \frac{G_F \ U_{us}} {\sqrt{2}}
\left [ J^\dagger_{\mu L} \ J_{\mu L} + \ \eta e^{i \beta} J^\dagger_{\mu R} \ J_{\mu
R}
\right ]
\end{equation}
where $\eta = m^2_{WL} /m^2_{WR}$ and $\beta$ is the relevant CPV phase.  Then
$H_{p.c.}$ and $H_{p.v.}$ have overall phases $(1 + i \eta \beta)$ and
$(1-i \eta  \beta)$ respectively.  To account for the observed CPV in
K-decay, $\eta \beta$
has to be of order $4.5.10^{-5}$.  In this model, $\epsilon'/\epsilon = 0$
and there are no rate asymmetries in hyperon decays but the asymmetries
 A and B are not zero and e.g. A goes as
$2 \eta \beta \sin (\delta_s - \delta_p)$.  In the class of models where
$W_L - W_R$ mixing is allowed, the hyperon asymmetries can be enhanced, and also
$\epsilon'/\epsilon$ is not zero in general\cite{chang} (see Table 1).

In MSSM (Minimum Supersymmetric Standard Model) there are new CP
violating phases and potentially new contributions to many observables.
Until recently the conventional thinking was that the most relevant
phase was the one in the squark LL mass terms:
\begin{equation}
m^2_{12} \ \overline{\tilde d}_L \ \tilde{s}_L
\end{equation}
and well constrained by $\epsilon$ so that the contribution to
$\epsilon'/\epsilon$ would be less than $2.10^{-4}$ (similarly for
hyperon decays).  The new wisdom, painfully learnt after the new results
on $\epsilon'/\epsilon,$ is that this is not the whole story. There are
several ways in which supersymmetric contributions can arise for $K$ and
hyperon decays.

One example is the lack of degeneracy of $\tilde{d}_R$ and $\tilde{u}_R$
masses\cite{kagan}.  This gives rise to I-spin breaking and in turn can enhance
$ImA_2$ and contribute to $\epsilon'/\epsilon$ at a level of
$10^{-3}$.  
For hyperon decays this would lead to a mild enhancement of
rate asymmetries but would have no effect on the asymmetries A being
probed by E871.

Another possibility is the existence of phases in the L-R squark mass
terms\cite{masiero}.  The effect of these on the s-d gluon dipole operator can be
parameterised as:
\begin{equation}
\left \{ a_{LR} \ \bar{d}_L \ \lambda^a \ \sigma_{\mu \nu} \ s_R \ + 
a_{RL} \ \bar{d}_R \ \lambda^a \ \sigma_{\mu \nu} \ s_L \right \} G_{\mu \nu}^a
 + h.c.
\end{equation}
In terms of $a_{LR}$ and $a_{RL}$, $\epsilon'/\epsilon$ and $A(\Lambda)$
 can be written as\cite{he}:
\begin{eqnarray}
\epsilon'/\epsilon \ \ \alpha  & Im (a_{LR} -a_{RL})  \\ \nonumber
A(\Lambda)         \ \  \alpha & Im(0.2 \ a_{LR} + 2.6 a_{RL}).
\end{eqnarray}
The figure shows the range of $A (\Lambda)$ for various allowed values
of $a_{LR}$ and $a_{RL}$.  Note that $a_{RL}$ can yield values for
$A(\Lambda)$ as large as $10^{-3}$ easily probed by E871.  This operator
is also enhanced in models where CP violation arises thru the exchange
of charged scalars such as the Weinberg model\cite{wein}.  

\begin{table}
\caption{Expectations for Hyperon CPV Asymmetries.}
\begin{center}
\begin{tabular}{|lccccc|} \hline
                   &   \mbox{SM}   & \mbox{2-Higgs}    &   \mbox{FCNC}  &  L-R-S  &  L-R-S  \\ 
                   &               &    	       &  \mbox{Superweak}  & (1)   &  (2) \\ 
$\Delta_\Lambda$   &  $10^{-6}$    &   $10^{-5}$       &        0
                                                                            &   0    &  0  \\
$A_\Lambda$   &  $ -4.10^{-5}$    &   $-2.10^{-5}$  &      0  & $-10^5$ &   $6.10^{-4}$  \\
$B_\Lambda$   &$10^{-4}$             &   $2.10^{-3}$     &  0    &
                   $7.10^{-4}$  & -  \\
$\Delta_\Xi$   & 0  & 0   &  0   &  0  &  0  \\
$A_\Xi$   &  $-4.10^{-6}$    &   $-3.10^{-4}$  & 0    & $2.10^{-5}$     & $10^{-4}$  \\
$B_\Xi$   &  $10^{-3}$    &   $ 4.10^{-3}$  & 0    & $3.10^{-4}$     &   -
                   \\ \hline
\end{tabular}
\end{center}
\end{table}

\section{Experiments}

There have been several proposals to measure hyperon decay asymmetries in
$\bar{p} p \ra \bar{\Lambda} \Lambda, \ \bar{p}{p} \ra \overline{\Xi} \Xi$ and in
$e^+e^- \ra J/\psi \ra \Lambda \overline{\Lambda}$ but none of these
were approved \cite{hamman}.  The only approved and on-going experiment is E871 at
Fermilab.  In this experiment $\Xi^-$ and $\overline{\Xi}^+$ are produced
and the angular distribution of $\Xi^-  \ra \Lambda \pi^- \ra p \pi^- \pi^-$ and
$\overline{\Xi}^+$ compared.  This experiment effectively measures $A_\Lambda +
A_\Xi$ and will be described in detail by Kam-Biu Luk \cite{luk}.
To summarize the implications for the measurement of $A_\Lambda + A_\Xi$
by E871: the SM expectation is about $-4.10^{-5}$ with a factor of two
uncertainty; if new physics should contribute it could be as large as 
$10^{-3}$. A measurement by E871 at the $10^{-4}$ level, therefore, will already 
be a strong discriminant.  Eventually, it will be important to know
$A_\Lambda$ and $A_\Xi$ separately and the old proposals\cite{hamman}
should be revived.

\section{$\epsilon'/\epsilon$ and Hyperon Decay Asymmetries}

It might seem that now that $\epsilon'/\epsilon$ has been measured and
direct CP violation in $\Delta S=1$ channel been observed, a study of CP
violation in hyperon decays is unnecessary and no new information will
be obtained. Why is it worthwhile measuring
another $\Delta S=1$ process like hyperon decay?  The point is that there
are important differences and the two are not at all identical.
First, there are important differences in the matrix elements. Hyperon
matrix elements do not have the kind of large cancellations that plague
the kaon matrix elements. The hadronic uncertainties are present for both,
but are different. Next, a very important difference is the fact
that the K $\ra \pi \pi$ decay (and hence $\epsilon'$) is only sensitive
to CP violation in the parity violating amplitude and cannot yield any
information on parity conserving amplitudes. Hyperon decays, by contrast, are
sensitive to both. Thus, $\epsilon'/\epsilon$ and hyperon decay
CP asymmetries are different and complimentary. The hyperon decay
measurements are as important and significant as $\epsilon'/\epsilon$.

\section*{Conclusion}

The searches for direct CPV are being pursued in many channels:  
$\Lambda \ra N {\pi}$, B decays and D decays.  Any observation of
a signal would be the first outside of $K^0 - \overline{K}^0$ system and
would be complimentary to the measurement of $\epsilon'/\epsilon$. 
This will constitute one more step in our bid to
confirm or demolish the Standard Kobayashi-Maskawa description of CP 
violation.

Hyperon decays offer a rich variety of CP violating observables, each
with different sensitivity to various sources of CP violation.  For
example, $\Delta_\Lambda$ is mostly sensitive to parity
violating amplitudes, $\Delta_{\Sigma +}$ is sensitive only
to parity conserving amplitudes, $A$ is sensitive to both etc.  The size
of expected signals vary inversely with the ease of making measurements,
i.e. $\Delta < A < B$. Probably
because of that, the
number of experimental proposals is rather small so far.  The one
on-going experiment Fermilab E871 can probe $A$ to a level of $10^{-4}$
which is already in an interesting range.  In addition to more
experiments, this subject sorely needs more
attention devoted to calculating the matrix elements more reliably.
   
\section*{Acknowledgment}

I am grateful to my collaborators Alakabha Datta,
Xiao-Gang He, Hitoshi Murayama, Pat O'Donnell, 
German Valencia and John Donoghue and to
members of the E871 collaboration for many discussions. 
The hospitality of Hai-Yang Cheng, George Hou and their
colleagues and staff was memorable and the atmosphere of the conference was 
most stimulating. This work is supported in part by USDOE under Grant \#DE-
FG-03-94ER40833.

\begin{figure}[h]
  \centerline{\psfig{file=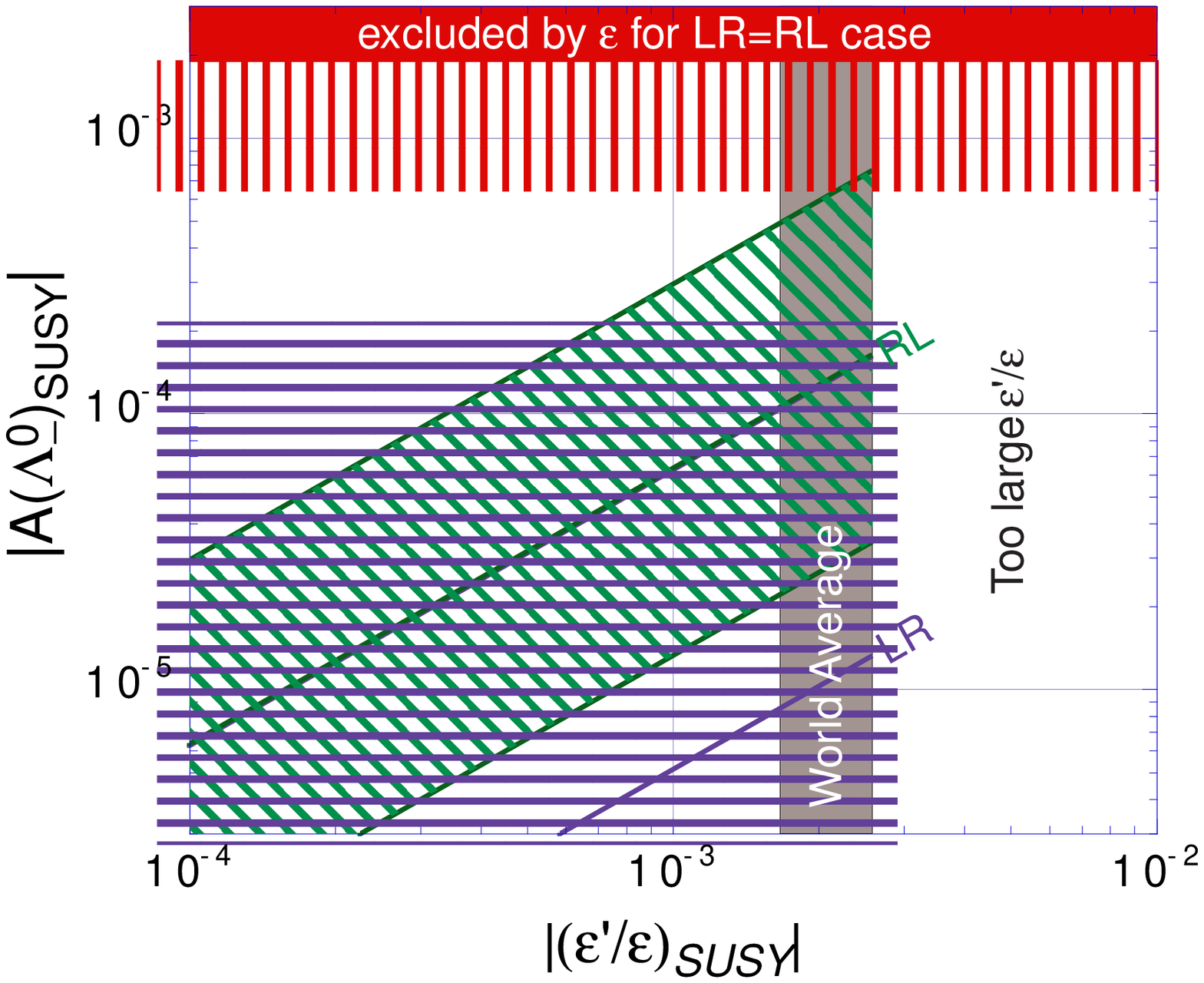,width=0.75\textwidth}} 
  \caption{The allowed regions on $(|(\epsilon'/\epsilon)_{SUSY}|,$ 
    $|A(\Lambda^{0}_-)_{SUSY}|)$ parameter space for three cases: a)
    only $Im(a_{LR})$ contribution, which is the
    conservative case (hatched horizontally), b) only ${\rm
      Im}(a_{RL})$ contribution (hatched diagonally), and
    c) $Im(a_{LR})=Im(a_{RL})$
    case which does not contribute to $\epsilon'$ and can give a large
    $|A(\Lambda^{0}_{-})|$ below the shaded region (or vertically
    hatched region for the central values of the matrix elements).  The
    last case is motivated by the relation $\lambda =
    \sqrt{m_{d}/m_{s}}$.  The vertical shaded band is the world
    average of $\epsilon'/\epsilon$.  The region to
    the right of the band is therefore not allowed.}
        \label{figure}
\end{figure}
\end{document}